%% file: EconPaper_revised.tex
\documentclass{article}

\usepackage[english]{babel}

\usepackage[letterpaper,top=2cm,bottom=2cm,left=3cm,right=3cm,marginparwidth=1.75cm]{geometry}

\usepackage{amsmath}
\usepackage{amsfonts}
\usepackage{tikz}
\usepackage{graphicx}
\usetikzlibrary{decorations.pathreplacing}
\usepackage{float}
\usepackage{placeins}
\usepackage{graphicx}
\usepackage{subcaption}
\usepackage[colorlinks=true, allcolors=blue]{hyperref}
\usepackage{booktabs}
\title{Measuring Growth and Convergence at the Urban-Scale}
\author{Isaak Mengesha$^\dagger$ and Debraj Roy$^\dagger$}
\date{
	$^\dagger$University of Amsterdam \\ %\texttt{\{auth1, auth3\}@org1.edu}\\%
	\today
}

\begin{document}
\maketitle

\begin{abstract}
Global inequality has shifted inward, with rising dispersion increasingly occurring within countries rather than between them. Using 8,790 newly harmonised Functional Urban Areas (FUAs) (micro-founded labour-market regions encompassing 3.9 billion people and representing approximately 80\% of global GDP) we show that national aggregates systematically, and increasingly, misrepresent the dynamics of growth, convergence, and structural change. Holding the underlying nighttime-lights GDP raster (1992--2019) fixed while varying the unit of aggregation (ADM0--ADM3, FUA), we isolate the contribution of the  Modifiable Areal Unit Problem directly in the growth literature. Three results follow. First, where inequality is located is unit-dependent: FUAs recover the most stable income–inequality relationship, corroborated (suggestively, $p\approx0.10$) by independent wealth data. Second, estimated $\beta$-convergence is scale-sensitive, and FUAs exhibit a discrete jump in convergence strength relative to administrative units of comparable population. Third, we find no poverty trap at the urban scale: expected growth remains positive throughout the income distribution, while the middle-income acceleration flattens over time. The cross-country convergence debate has been conducted on national aggregates, yet nations are political containers rather than economic units, and measured relationships, including the convergence coefficient, depend on the spatial unit of analysis.
\end{abstract}

\section{Introduction}

Global inequality has undergone a fundamental geographic reorientation. While between-country income gaps remain substantial, within-country dispersion is rising across the world economy \cite{un_inequality}. Therefore, studying growth at the national level misses where heterogeneity is increasingly located. Yet other spatial disaggregation -- i.e. sub-national administrative boundaries -- introduces systematic distortions through the Modifiable Areal Unit Problem (MAUP) \cite{openshaw1984modifiable}. Statistical relationships depend on the scale and zoning of underlying spatial units. Administrative boundaries aggregate  historical circumstances rather than economic integration, obscuring the mechanisms generating divergence, convergence, and structural adjustment. Evidence for multiple steady states at sub-national scales is robust, yet global convergence analysis remains constrained by administratively defined regions rather than economically meaningful spatial units \cite{durlauf1995multiple,bartkowska2012regional,gennaioli2014growth}. When development constraints are locally embedded, national convergence strategies based on place-neutral policy reasoning are structurally misaligned with the geography of growth. \cite{barca2012case}.\\

Functional Urban Areas (FUAs) provide a theoretically grounded alternative. Delineated using population grids and commuting-flow thresholds, FUAs approximate integrated labour-market regions rather than political jurisdictions \cite{dijkstra2019eu}. This matters because knowledge-intensive activities concentrate in cities \cite{BJFP+20}, FUAs capture actual adjustment margins to labour-demand shocks \cite{tolbert1996us}, and approximately 3.9 billion people live in 8,790 FUAs worldwide \cite{moreno2021metropolitan}. As a result, FUAs provide the appropriate spatial scale for studying growth heterogeneity as an outcome of endogenous adjustment.\\

Observed at the FUA level, urban systems display systematic development heterogeneity. Larger cities enjoy persistent productivity premia \cite{ahrend2014makes}, while mature urban systems exhibit diminishing population-growth advantages \cite{musso2025large}. These regularities imply distinct urban growth patterns that are mechanically blurred by national aggregates.\\

This paper asks how the choice of spatial aggregation shapes what we measure as convergence, and where we locate inequality in the world economy. It makes three contributions. First, holding the GDP data fixed, the unit of analysis materially changes the estimated convergence coefficient---the MAUP operating directly in the growth literature. Second, the location of inequality (between- versus within-country) is unit-dependent, and FUAs recover a systematic relationship that administrative units obscure. Third, at the FUA scale there is no evidence of a poverty trap, and the middle-income acceleration is present but flattening. We are explicit about scope: the paper documents that spatial scale matters and characterises the urban income--growth relationship, but does not claim to identify the mechanism generating persistent urban income differences.

\section{Related Literature}

Two bodies of research inform this paper's approach to urban growth dynamics: the convergence debate and the conditions under which economies catch up, fall behind, or cluster into distinct steady states; and the use of functional urban areas as units of economic analysis. A common theoretical concern cuts across both. New Economic Geography demonstrates that increasing returns and transport costs generate multiple spatial equilibria rather than a unique steady state \cite{krugman1991increasing}, and that economic outcomes depend on market access and spatial integration rather than administratively defined boundaries \cite{fujita2001spatial}. Aggregating heterogeneous spatial equilibria at the national level therefore mechanically obscures growth and convergence dynamics. Spatial scale is not a neutral measurement choice but a theoretically consequential one---an insight that motivates our effort to operationalise convergence analysis at the scale of functional urban areas.

\subsection*{Convergence and the conditions for catch-up}

Recent work has reignited debate over whether poorer economies are catching up to richer ones. \cite{patel2021new} documented unconditional convergence at the country level since the 1990s, yet the authors themselves have since noted a post-2010 reversal \cite{patel2025convergence}, corroborated by others \cite{olaberria2023reversal}. This instability is puzzling on its own terms, but becomes more so when paired with evidence that between-country convergence has coincided with rising inequality within countries \cite{alvaredo2018elephant}. Together, this puts the conventional approach of measuring convergence at the national level into question.\\

To situate this concern, consider the theoretical foundations. The Solow model predicts convergence to a common steady state \cite{solow1956contribution}, and early evidence among relatively homogeneous samples---OECD economies, US states---appeared consistent with this prediction \cite{barro1992convergence}. Globally, however, unconditional convergence does not agree with data. The resolution proposed by \cite{mankiw1992contribution} is conditional convergence: economies converge not to a universal equilibrium but to heterogeneous steady states determined by structural characteristics---savings rates, population growth, human capital, institutions. Convergence then operates within groups sharing similar fundamentals; across groups, income gaps persist. Models incorporating nonconvexities or complementarities go further, generating multiple locally stable equilibria \cite{azariadis1990threshold} and the empirical implication of convergence clubs: catch-up within clusters, divergence between them \cite{durlauf1995multiple}.\\

\subsection*{A natural and functional unit of analysis}

Agglomeration economies arise from localized mechanisms---learning, matching, and sharing---that operate within integrated labour markets and dense interaction zones \cite{duranton2004micro, fujita2002agglomeration}. Administrative regions do not generally coincide with the spatial extent of these mechanisms. Subnational units such as NUTS regions, provinces, and districts remain artefacts of political history rather than economic integration, splitting integrated labour markets or aggregating distinct ones and thereby introducing measurement error that biases growth estimates \cite{anselin2022spatial}. Functional Urban Areas offer a principled alternative. Because FUAs are delineated by commuting flows, they capture the spatial extent of daily interaction---the boundary within which knowledge spillovers, labour pooling, and input sharing actually operate \cite{moreno2021metropolitan}.\\

The broader agglomeration literature, while documenting systematic productivity and growth differences across cities \cite{combes2015empirics, rosenthal2004evidence}, typically takes the spatial unit of observation as given rather than examining how aggregation alters inferred growth laws or convergence estimates \cite{glaeser1992growth}. Modern spatial economics emphasises quantitative multi-region equilibrium analysis in which outcomes depend on the structure of spatial interaction across locations \cite{redding2017quantitative}, but empirical convergence analysis remains largely conducted at administrative scales. This paper integrates the spatial logic of economic geography into convergence empirics by changing the unit of analysis. Whether the patterns documented in cross-country research---conditional convergence, multiple equilibria, convergence clubs---replicate, intensify, or reverse when examined across functionally defined cities remains an open question. Limitations of the FUA approach remain---commuting patterns may lag economic restructuring, and coverage is uneven in low-income settings---but FUAs nonetheless represent the best available approximation to integrated urban economies.

\begin{figure}[t] 
    \centering 
    \input{maup_partition.tex} 
    \caption{Two partitions of the identical nighttime-lights raster. Administrative boundaries (purple) cut across the labor-market basin, mixing basin and hinterland within each unit; the functional boundary (green) encloses it. Estimating Equation~\ref{eq:solow} on these units yields attenuated convergence under the political partition and recovers the signal under the functional one---a zoning effect at fixed scale.} 
    \label{fig:maup-partition} 
\end{figure}

\section{Methods}
\subsection*{Estimating Convergence}

The cross-country growth regression derived from the Solow model relates the average growth rate of output per capita to its initial level. We estimate the nonlinear specification of \cite{patel2021new},
\begin{equation}
    \frac{1}{s}\ln\!\left(\frac{y_{i,t+s}}{y_{i,t}}\right) = \alpha - \frac{1-e^{-\beta s}}{s}\ln y_{i,t} + \varepsilon
    \label{eq:solow}
\end{equation}
where $y_{i,t}$ is initial output per capita, $s$ is the horizon over which growth is measured, and $\alpha$ absorbs common steady-state determinants such as investment rates and population growth. A positive coefficient $\beta > 0$ indicates convergence: poorer economies grow faster due to diminishing returns to capital. The parameter $\beta$ is estimated by nonlinear least squares following \cite{patel2021new}. This specification assumes all economies share a common production function and converge toward a common steady state. In our subsequent analysis we estimate $\beta$ for statements of convergence and divergence.

A single $\beta$ estimated by pooling heterogeneous spatial units can be misleading: it can inherit within-group convergence while masking across-group divergence \cite{durlauf1995multiple}. This is precisely why the choice of spatial unit---how units are drawn and at what scale---is not innocuous for the measured convergence coefficient.

\subsection{Data and Measurement}

\begin{figure}[H]
    \centering
    \includegraphics[width=0.7\linewidth]{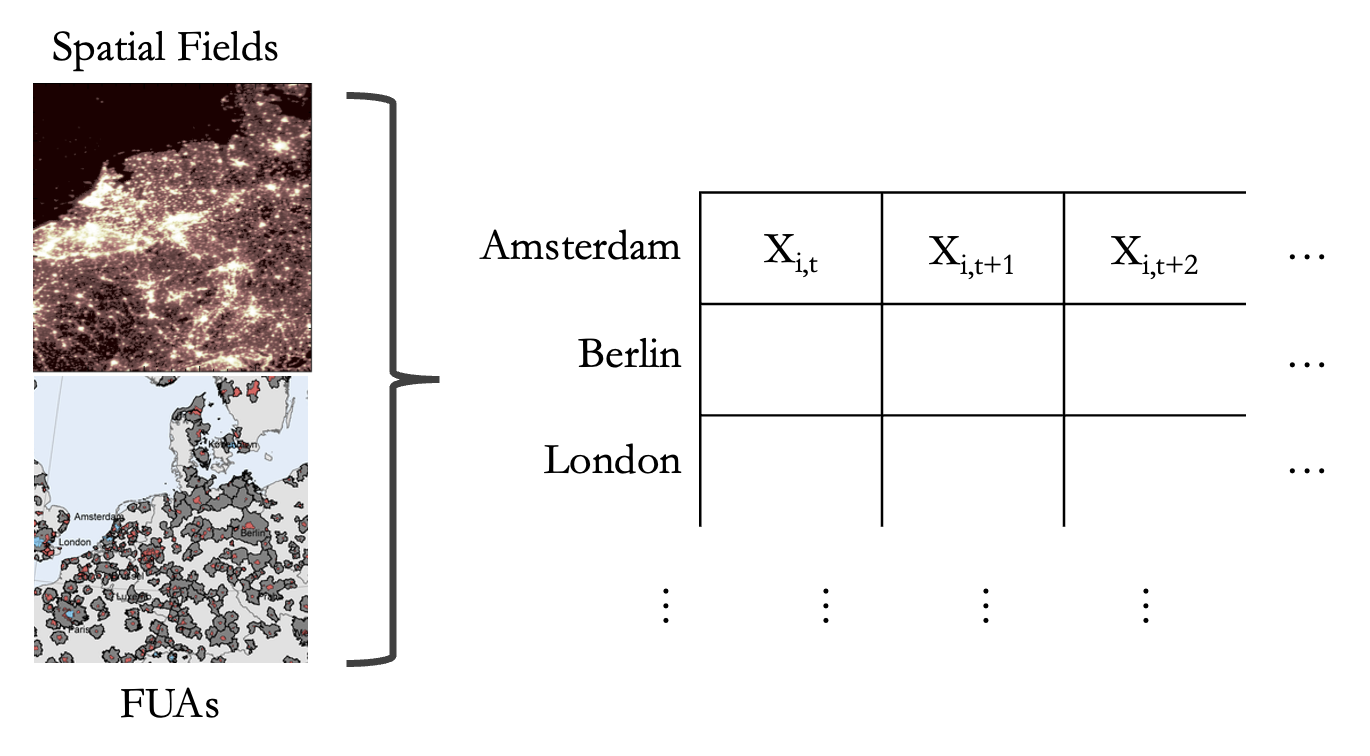}
    \caption{We utilize FUAs as aggregating unit over various spatial fields, including GDP, Population, and the Relative Wealth Index (RWI). With this we construct the timeseries of observation for each FUA.}
    \label{fig:schema}
\end{figure}

The analysis combines globally harmonised functional urban areas with high–resolution economic data to construct a panel of urban economies from 1992–2019. The objective is to measure GDP dynamics at the level of microfounded labour–market regions. Figure~\ref{fig:schema} provides a schematic overview of the aggregation procedure.

\vspace{0.5em}
\noindent\textbf{Functional Urban Areas.}
Urban boundaries follow the \textit{GHS–FUA 2023} delineation, which identifies 8{,}790 functional labour–market regions using globally consistent population grids and commuting thresholds \cite{ghs_fua_2023}. All FUAs are retained. Each 1km cell from the underlying GHSL raster is assigned to the FUA whose polygon contains the majority of its area; this avoids multi–assignment and reflects the empirical indivisibility of urban labour–market basins. FUAs with missing years in the outcome series are omitted ($<5\%$ of FUAs), but no further sample restrictions are imposed.

\vspace{0.5em}
\noindent\textbf{GDP and Population Fields.}
Economic activity is measured using the global 1km GDP panel of \cite{Chen2022}, which provides harmonised yearly estimates for 1992–2019 derived from calibrated nighttime‐lights (NTL) data. Population is taken from the GHSL population layers, linearly interpolated to yearly values, so that per-capita levels and growth reflect year-varying population. For each FUA $i$ and year $t$, total GDP is obtained by summing cell–level estimates within the FUA boundary, and GDP per capita is constructed as
\[
y_{i,t}=\frac{\text{GDP}_{i,t}}{\text{Population}_{i,t}}.
\]
Growth is defined as the annual percentage change,
\[
g_{i,t}=\frac{y_{i,t+1}-y_{i,t}}{y_{i,t}}.
\]
No additional scaling, PPP adjustment, or reconciliation with national accounts is applied; all values are taken directly from the harmonised dataset. As pointed out by related literature there exist several limitations to estimating GDP from NTL. Some point at the instability at the regional level \cite{bickenbach2016night}, or the subannual level \cite{perez2021night}. Finally there is evidence for saturation bias, in large and more developed regions \cite{mellander2015night}.

\noindent\textbf{Construction of the FUA Panel.}
The resulting dataset consists of yearly observations
\[
X_{i,t} = \bigl( y_{i,t}, g_{i,t}, \text{GDP}_{i,t}, \text{Pop}_{i,t} \bigr),
\]
allocated to each FUA through the time–dependent spatial fields described above. Grid–level GDP enters additively and population varies yearly through the interpolated GHSL layers. This produces a coherent panel of urban economies suitable for analysing transition dynamics and convergence patterns. The schematic in Figure~\ref{fig:schema} summarises the mapping from raster data to FUA–level aggregates and the subsequent construction of the variables used in the empirical analysis.

\newpage
\FloatBarrier
\section{Empirical Results}

The empirical analysis proceeds in three steps. We first document how the global distribution of production changes over time and -- more importantly -- over spatial aggregations for both production and wealth. Next we investigate global growth and convergence across spatial aggregations and contrast them with the broader literature. Finally we examine the urban law of motion and assess the evidence for poverty traps.

\subsection{Where is inequality located in the world economy?}

We locate inequality in the world economy by decomposing the cross-sectional variance of GDP per capita across alternative spatial aggregations. For each year from 1992 to 2019, we decompose scale-level GDP per capita variance into between-country and within-country components, allowing us to assess how the geography of inequality depends on the unit of observation.\\

Two robust patterns emerge. First, between-country inequality dominates at coarse spatial scales but declines monotonically over time. Second, the within-country component rises steadily, indicating a persistent inward shift of inequality. This pattern holds throughout the sample period and is consistent with global convergence accompanied by increasing internal dispersion. However, the level at which within-country inequality appears dominant depends strongly on spatial aggregation. At the ADM1 level, most variance is mechanically attributed to between-country differences. Finer administrative units increase the within-country share, but do not eliminate sensitivity to boundary choice. When inequality is measured at the level of Functional Urban Areas (FUAs), a clear separation re-emerges: more than 75\% of total variance is again accounted for by between-country differences. Measured at the scale of integrated labour markets, cross-country inequality is therefore substantially more stable. We note that nighttime-lights-based GDP underrepresents rural and agricultural activity, and that space-filling administrative units may introduce additional aggregation error.\\

% NOTE (Isaak): the abstract frames inequality as shifting "inward",
% while this paragraph reports that FUAs attribute >75% of variance to
% the between-country component. Worth a sentence reconciling the two
% readings -- e.g. FUAs recover a *stable* between/within split where
% administrative units are boundary-dependent -- so the section and the
% abstract tell the same story.

To verify that these results are not specific to GDP measurement, we run a similar analysis using an independent wealth metric. We employ the Relative Wealth Index (RWI), an asset-based household wealth measure derived from DHS and related surveys \cite{rwi_hdx}. For countries with at least twelve FUAs, we decompose within-country wealth variance across spatial scales and regress within-unit inequality on national GDP per capita. Spatial scale again proves decisive. At administrative levels (ADM1–ADM3), there is no systematic relationship between income and internal inequality. At the FUA level, a positive association emerges: richer countries exhibit greater wealth dispersion within their urban systems. While only suggestive ($r^2 = 0.146$, $p = 0.097$), this relationship disappears under administrative aggregation. Within-FUA inequality accounts for roughly 85–90\% of national RWI variance, compared to less than 50\% at the ADM2 level, where estimates are also substantially more volatile. Across all spatial levels, within-unit inequality increases with GDP per capita, but only FUAs recover a systematic relationship, though the association is statistically weak.\\

These results have three implications. First, the Modifiable Areal Unit Problem is not merely a statistical inconvenience: administrative boundaries systematically distort the spatial allocation of inequality by misaligning economic integration with political geography. Second, the findings provide a spatial interpretation of the dynamics underlying Milanović’s elephant curve: as between-country gaps narrow, inequality rises within countries, concentrated in urban labour markets rather than administrative regions \cite{milanovic2016global}. Third, the evidence contradicts a subnational Kuznets hypothesis. Internal spatial inequality does not decline with development; instead, richer urban systems exhibit greater internal differentiation, consistent with agglomeration models in which productivity gains increasingly concentrate in dense labour markets \cite{combes2015empirics}.

\begin{figure}[H]
    \centering

    \includegraphics[width=0.95\linewidth]{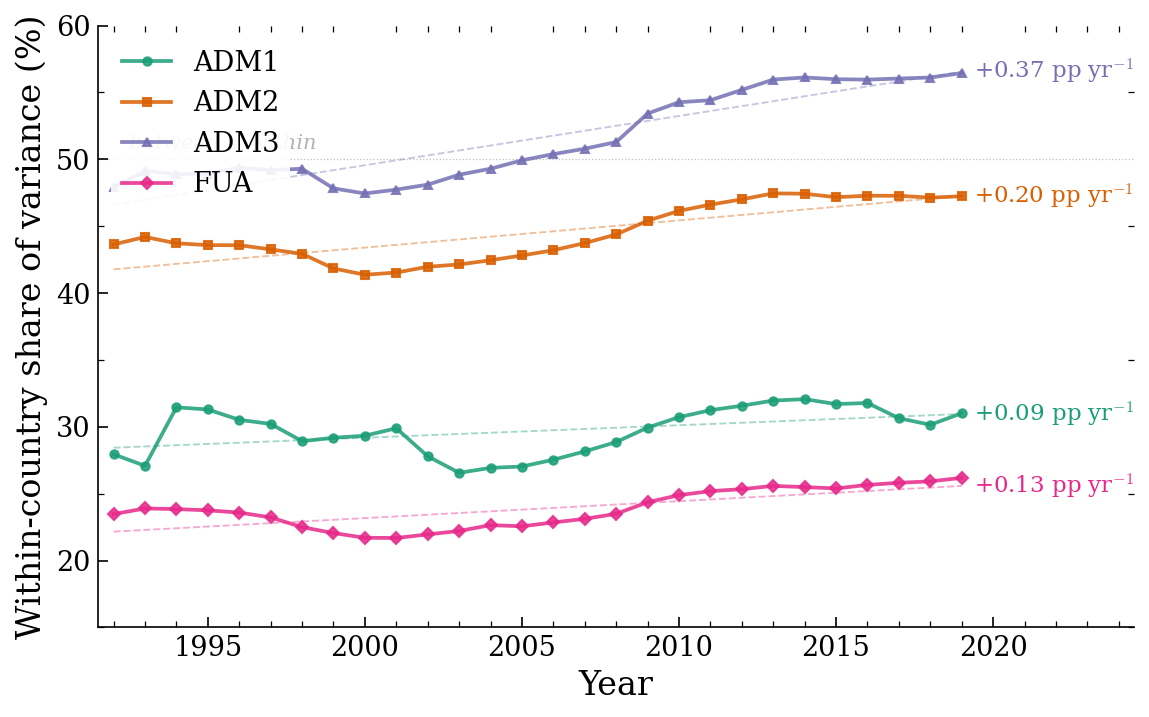}
    \caption{\textbf{Spatial inequality structure: functional vs. administrative geography.} Decomposing GDP per capita variance into between-country (purple) and within-country (red) components (1992--2019) for different ADM levels. Qualitatively similar behaviour at different intensities. }
    \label{fig:inequality}
\end{figure}

% From a policy perspective, these patterns imply that place-based interventions targeted at administrative regions are structurally misaligned with the geography of inequality. From a methodological perspective, they justify the use of FUAs as the relevant unit for subsequent analyses of convergence and capability accumulation. Functional urban areas recover relationships that disappear under alternative spatial schemes, revealing where inequality is generated and how it evolves as the world economy develops.

\subsection{Is convergence sensitive to scale?}

\begin{figure}[H]
    \centering
    \includegraphics[width=0.8\linewidth]{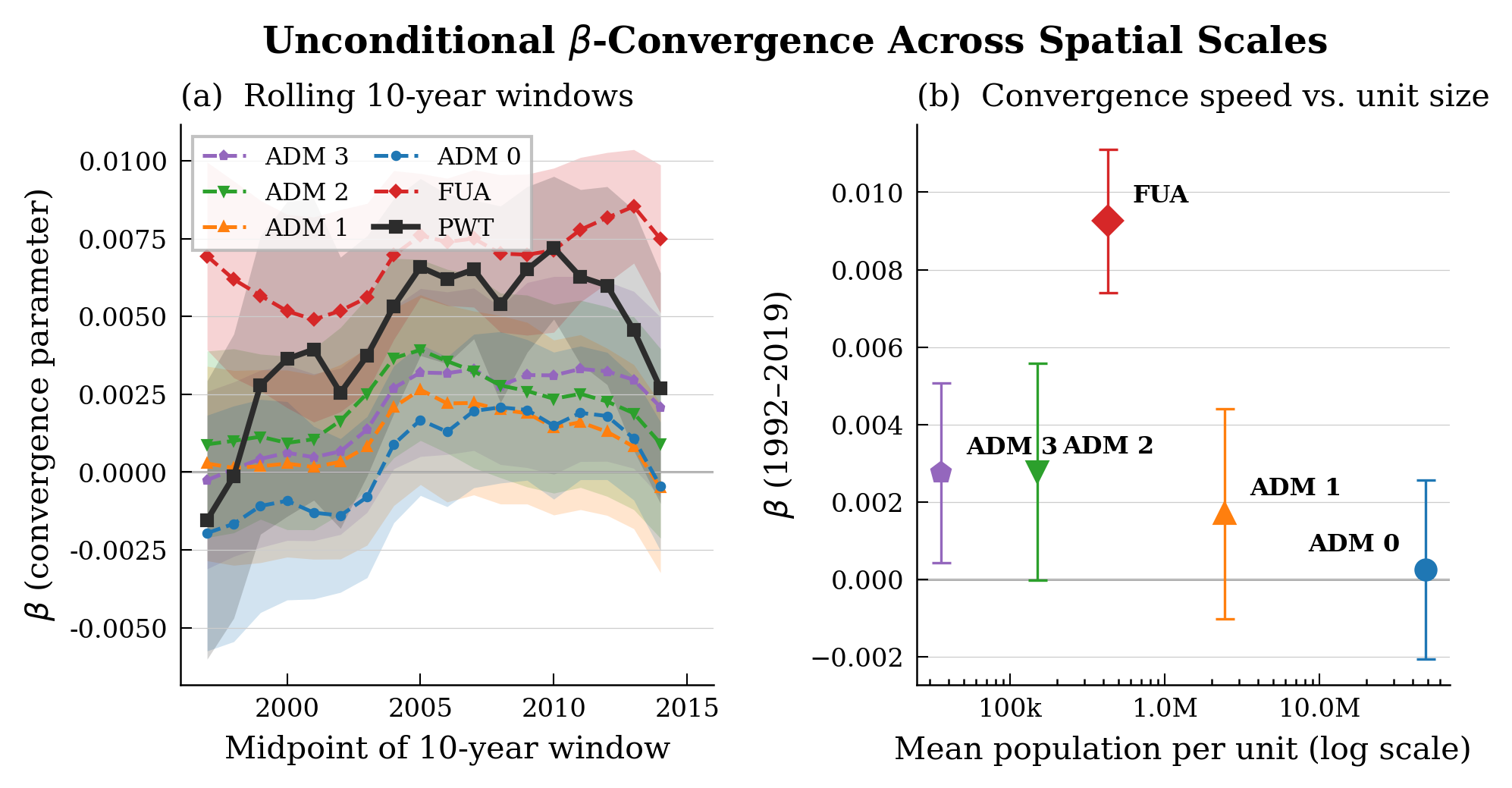}
    \caption{Estimated convergence is affected by the Modifiable Areal Unit Problem. Increasing urbanisation begs the question of which spatial disaggregation is appropriate. Whatever biases are introduced by NTL-based GDP estimates, they apply consistently across scales and therefore cannot account for the pronounced differences in convergence rates.}
    \label{fig:convergence}
\end{figure}

We assess whether measured convergence depends on the spatial unit of analysis by estimating $\beta$-convergence, following Equation~\ref{eq:solow}, across five spatial scales: countries (ADM0), administrative subdivisions (ADM1–ADM3), and Functional Urban Areas (FUAs). This allows us to isolate the role of spatial aggregation while holding the underlying GDP data fixed. Estimates derived from nighttime lights are noisy and biased \cite{mellander2015night,Chen2022}; every unit at every scale is built from the same raster, so aggregation reweights this error but cannot introduce new error. Composition does differ—FUAs concentrate dense, saturation-prone pixels—which the saturation check below addresses. To anchor the magnitudes, we benchmark country-level estimates against the Penn World Tables. As expected, NTL-based ADM0 estimates show weaker convergence than PWT, consistent with attenuation from measurement noise (see Fig \ref{fig:convergence}a). FUAs exhibit stronger convergence than both administrative units and PWT. One might worry this reflects saturation bias, which compresses high-income estimates and could mechanically generate faster apparent catch-up \cite{mellander2015night}. However, FUAs also exhibit the highest between-country variance share across all spatial scales (Figure 2); if saturation dominated, this share would be attenuated. This indicates signal recovery rather than artifact. Standard errors for cross-scale comparisons are computed by subsampling each scale to the ADM0 sample size ($B=1000$), so that differences in $\beta$ are not driven by differences in sample size across scales.

Cross-dataset discrepancies at the country level are well documented—PWT, Maddison, and WDI differ substantially in levels, though long-run trends broadly align \cite{patel2021new,patel2025convergence}. Our contribution is to show that, even holding the dataset fixed, the unit of analysis materially affects measured convergence. This is the Modifiable Areal Unit Problem operating directly in the growth literature. Figure~\ref{fig:convergence}b illustrates these patterns. Rolling ten-year estimates show that convergence measured on administrative units clusters near zero throughout the sample period, while FUAs display consistently stronger convergence. Plotting convergence speed against mean population per unit reveals a saturation pattern: convergence strengthens as administrative units become smaller, then plateaus. FUAs, despite having population sizes comparable to ADM2 regions, exhibit a discrete jump in convergence strength.\\

A distinct concern is mean reversion, or Galton's fallacy \cite{friedman1992old}. Transient measurement noise in initial income enters the right-hand-side regressor and, with opposite sign, the growth outcome, mechanically biasing $\beta$ toward convergence; because the variance of this transient noise is larger for smaller spatial units, the bias predicts precisely the scale gradient we observe. Two features of the data argue against this reading. First, the bias declines with the estimation horizon (transient noise contributes on the order of $\sigma^2_\nu/s$ to measured growth) yet the FUA--ADM gap persists in the full-period estimates ($s=27$, Figure~4b), not only in the rolling ten-year windows. Second, noise-variance bias scales smoothly with unit size, whereas FUAs display a discrete jump in convergence strength at population sizes comparable to ADM2 units; a smooth noise gradient cannot produce a discontinuity. This argument is separate from the saturation-bias check above.\\

This discontinuity suggests that functional delineation captures economically meaningful integration that administrative boundaries fragment. The saturation pattern is consistent with FUAs approximating the spatial scale at which key adjustment mechanisms—labour mobility, knowledge spillovers, and capital reallocation—operate \cite{combes2015empirics}. Further disaggregation would fragment integrated labour markets, while coarser aggregation pools heterogeneous units, attenuating the convergence signal.

\subsection{Changes to the urban law of motion}

\begin{figure}[H]
    \centering
    \includegraphics[width=0.85\linewidth]{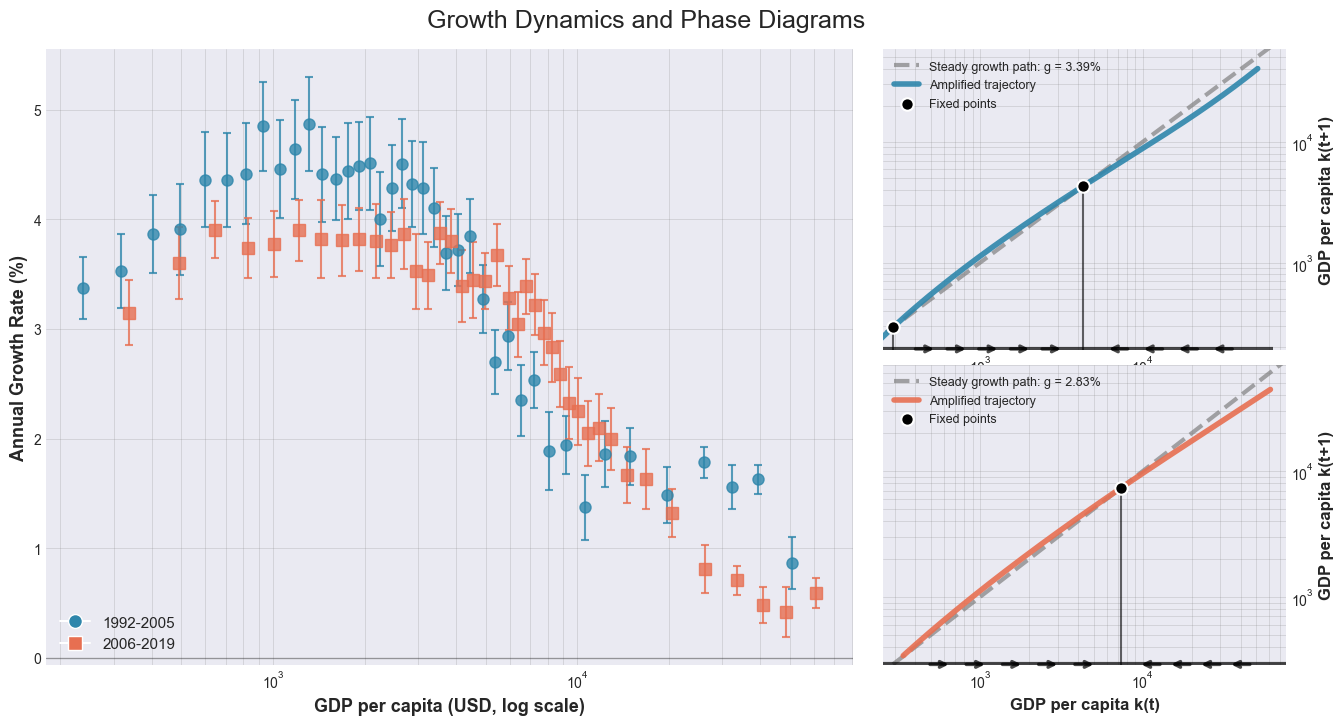}
    \caption{Flattening of the middle income trampoline in recent years. Here we construct the solow phase diagram, with a notable distinction. The diagonal represents the average growth trend, indicating that no FUA is in a development trap. Therefore, fixed points represent relative movement towards or away from each other. This pattern over different GDP levels indicates structural differences in growth across the income distribution. (Motion curve amplified)}
    \label{fig:solow}
\end{figure}

We examine how the relationship between income and growth at the urban level departs from the canonical Solow framework. Figure~\ref{fig:solow} (left) plots annual GDP per capita growth against initial income for all FUAs. Averaged across the full sample, the relationship is consistent with global convergence: poorer FUAs grow faster than richer ones. However, the conditional mean exhibits an inverted U-shape, indicative of a middle-income acceleration followed by slower growth at high income levels.\\

This pattern is not stable over time. In the later half of the sample period, the middle-income acceleration weakens substantially, and growth among the richest FUAs slows markedly. These shifts suggest that the urban growth process is not governed by a time-invariant law of motion, but evolves with the global development environment. Figure~\ref{fig:solow} (right) presents the corresponding Solow phase diagram. The 45-degree line represents average global growth; deviations from this line indicate relative convergence or divergence across FUAs. Standard poverty-trap models predict that the transition curve falls below the diagonal at low income levels, generating a locally stable low-income equilibrium \cite{azariadis1990threshold}. We find no evidence for such a mechanism. Expected growth remains positive for low-income FUAs throughout the income distribution.\\

The transition curve never intersects the diagonal from above at low income levels. There is therefore no locally stable poverty trap at the scale of functional urban areas. This finding rules out capital-threshold mechanisms as the primary explanation for persistent urban income differences, at least at currently observed capital levels \cite{solow1956contribution,azariadis1990threshold}. The implication is that urban income stratification cannot be accounted for within the standard Solow framework. Persistent divergence across cities must instead reflect heterogeneity in institutional environments or other factors shaping the urban growth process beyond factor accumulation alone.

\section{Discussion}

This paper has examined growth and convergence dynamics at various scales. We argued why the urban scale is a meaningful unit for analysing growth and convergence. The analysis yields a set of empirical regularities that challenge standard accounts of global development and reframe how we understand the geography of growth. We summarise these findings, note limitations, and outline directions for future research.\\

Three stylized facts emerge from the analysis. \textbf{(1)}~Global convergence rates are modulated by the spatial aggregation of recorded growth. Holding the underlying GDP data fixed, the estimated convergence coefficient changes systematically with the unit of analysis; administrative units cluster near zero while FUAs exhibit a discrete jump. Trends such as urbanization and other mechanisms point at FUAs as a meaningful unit of analysis. \textbf{(2)}~Classical poverty traps receive no empirical support at the urban scale. Low-income FUAs do not exhibit the negative or insufficient growth rates required to sustain a stable low-income equilibrium. The transition curve implies positive expected growth throughout the income distribution. Whatever generates persistent urban poverty, it is not the capital-threshold mechanism emphasised in canonical models—persistent divergence must arise from another factor rather than accumulation constraints. \textbf{(3)}~Inequality is increasingly a within-country and within-FUA phenomenon. The between-country share of global income variance has declined steadily since the early 1990s while the within-country share has risen -- across all spatial aggregations (this is a statement about the temporal trend in the GDP variance decomposition; in levels, between-country differences still account for the majority of variance at the FUA scale, Section 4.1). Moreover, the RWI results suggest that richer urban areas exhibit more wealth inequality. Inequality is not only shifting inward geographically; it is concentrating within the cities that drive national growth, highlighting the importance of asset- and productivity-based inequality metrics at the urban scale.\\

These findings are subject to limitations that bound the conclusions we can draw. The analysis is restricted to functional urban areas; rural regions—where structural transformation, market access, and agricultural productivity constraints may generate different dynamics—remain outside the dataset, and the absence of poverty traps at the urban scale does not imply their absence elsewhere. The evidence is descriptive rather than causal. Finally, GDP is proxied via calibrated nighttime-lights data, introducing measurement error that validation against subnational accounts could address.\\

NTL-based FUA GDP should be validated against subnational accounts where they exist, to bound the measurement error. The analysis should be extended to rural areas to test whether the scale-dependence of convergence is specific to cities. And the mechanisms generating the scale-dependence---labour mobility, capital reallocation, and the zoning versus scale effects of the MAUP---remain to be disentangled, which we leave open here. Functional urban areas reveal growth and convergence dynamics that national aggregates obscure: the spatial unit is a first-order choice, not a technicality.

\section{Acknowledgments}
The authors gratefully acknowledges institutional support from the University of Amsterdam.

\FloatBarrier
\bibliographystyle{alpha}
\bibliography{sample}
\FloatBarrier

\section{Supplementary Information}
\FloatBarrier
\begin{figure}[H]
    \centering
    \includegraphics[width=0.9\linewidth]{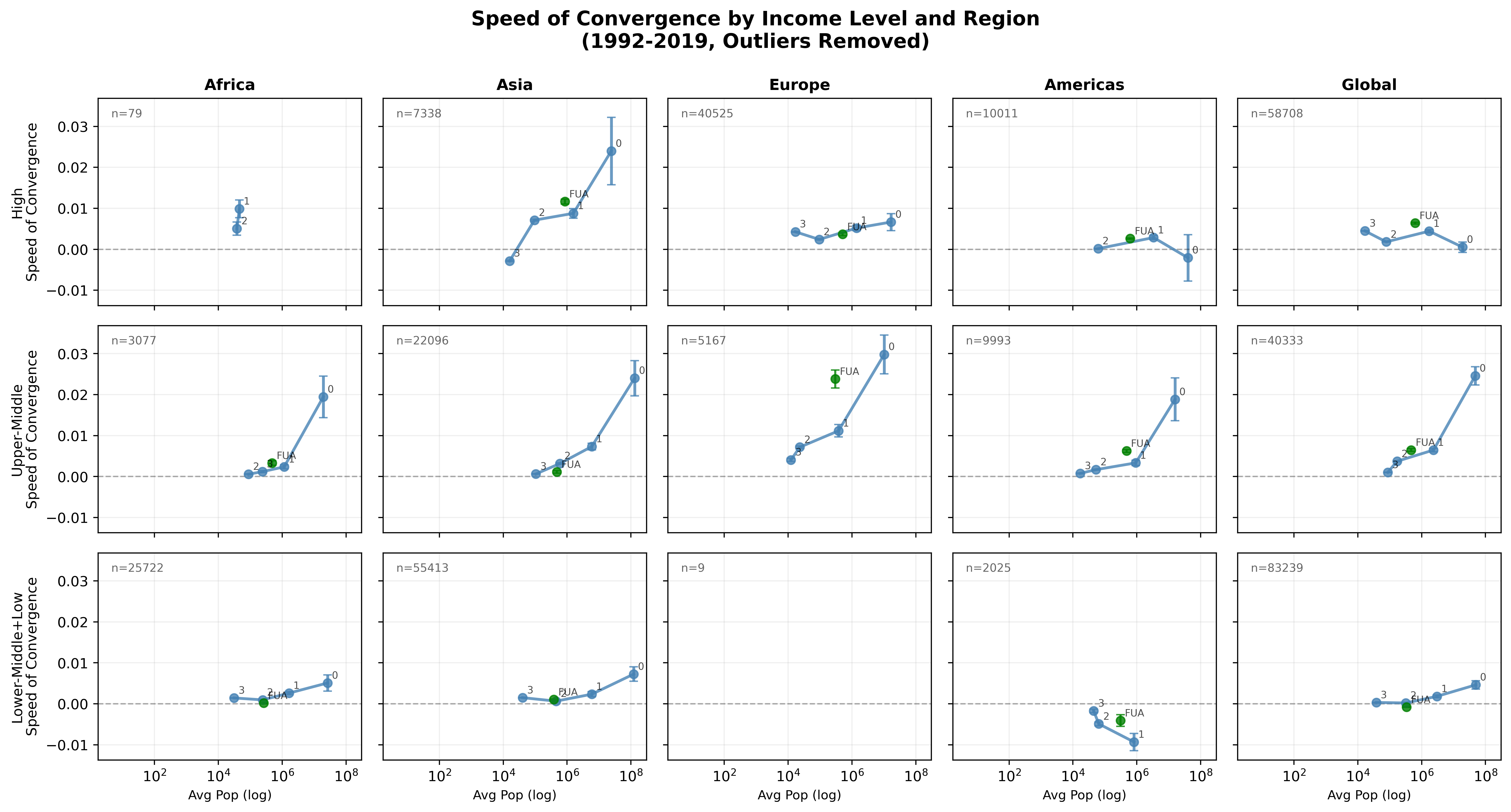}
    \caption{Convergence measured across different regions and different spatial scales.}
    \label{fig:conv_regions}
\end{figure}

\begin{figure}[p] \centering \includegraphics[width=\textwidth]{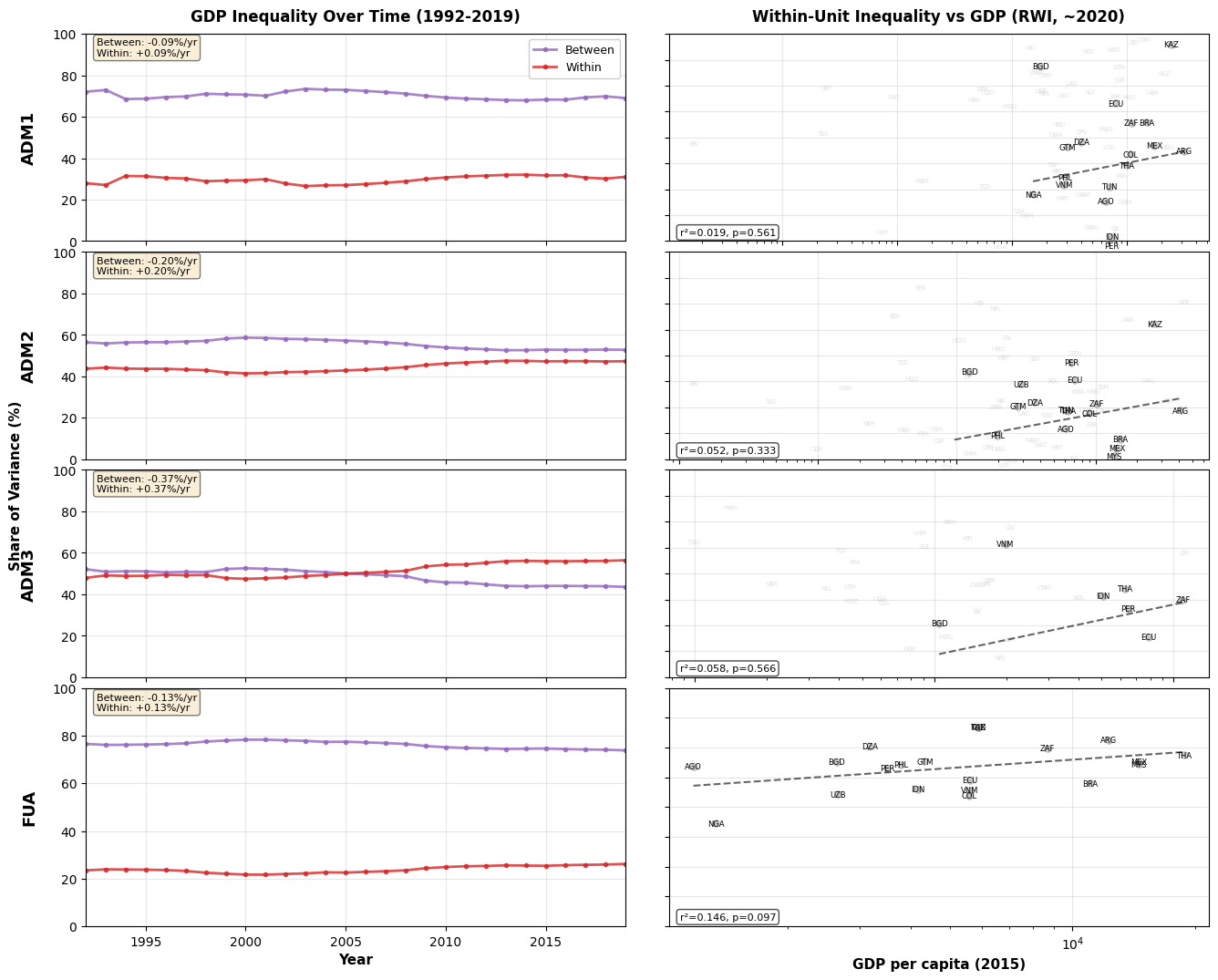} \caption{Spatial inequality structure across all aggregation levels (ADM1--ADM3, FUA). Left column: decomposition of GDP per capita variance into between-country and within-country shares, 1992--2019, with linear trends per year. Right column: within-unit RWI variance share against national GDP per capita ($\sim$2020) for countries with $\geq$12 units; dashed line shows the fitted relationship. Only FUAs yield a suggestive income--inequality relationship ($r^2=0.146$, $p=0.097$); administrative units show none ($p>0.33$).} \label{fig:inequality-adm-fua-full} \end{figure}

\end{document}

%% file: maup_partition.tex
\begin{tikzpicture}[x=0.016cm,y=0.016cm] \definecolor{ntl}{HTML}{EF9F27} \definecolor{adm}{HTML}{7F77DD} \definecolor{fuacol}{HTML}{1D9E75} \draw[gray!70,dashed,rounded corners=3pt,line width=0.4pt] (40,30) rectangle (320,180); \draw[gray!70,dashed,rounded corners=3pt,line width=0.4pt] (360,30) rectangle (640,180); \foreach \dx in {0,320}{\foreach \x/\y/\r/\o in {165/108/9/0.9,192/104/7/0.8,178/91/10/1.0,206/92/6/0.6,154/93/6/0.6,185/117/4/0.5,96/140/3/0.25,260/131/3/0.25,104/67/3/0.25,254/65/3/0.25,226/140/3/0.3,130/63/3/0.3}{\fill[ntl,opacity=\o] (\x+\dx,\y) circle[radius=\r];}} \draw[adm,line width=0.9pt] (172,180)--(165,138)--(181,102)--(166,59)--(173,30); \draw[adm,line width=0.9pt] (40,102)--(105,96)--(168,108)--(235,94)--(320,101); \draw[fuacol,dashed,line width=1pt] (501,102) ellipse[x radius=65,y radius=31]; \node[font=\large] at (180,8) {Political partition}; \node[font=\large] at (500,8) {Functional partition}; \end{tikzpicture}

%% file: sample.bib
@misc{friedman1992old,
  title={Do old fallacies ever die?},
  author={Friedman, Milton},
  journal={Journal of economic literature},
  pages={2129--2132},
  year={1992},
  publisher={JSTOR}
}

@article{openshaw1984modifiable,
  title={The modifiable areal unit problem},
  author={Openshaw, Stan},
  journal={Concepts and techniques in modern geography},
  year={1984},
  publisher={GeoBooks}
}

@article{solow1956contribution,
  author  = {Solow, Robert M.},
  title   = {A Contribution to the Theory of Economic Growth},
  journal = {Quarterly Journal of Economics},
  volume  = {70},
  number  = {1},
  pages   = {65--94},
  year    = {1956},
  doi     = {10.2307/1884513}
}

@article{mankiw1992contribution,
  author  = {Mankiw, N. Gregory and Romer, David and Weil, David N.},
  title   = {A Contribution to the Empirics of Economic Growth},
  journal = {Quarterly Journal of Economics},
  volume  = {107},
  number  = {2},
  pages   = {407--437},
  year    = {1992},
  doi     = {10.2307/2118477}
}

@article{barro1992convergence,
  author  = {Barro, Robert J. and Sala-i-Martin, Xavier},
  title   = {Convergence},
  journal = {Journal of Political Economy},
  volume  = {100},
  number  = {2},
  pages   = {223--251},
  year    = {1992},
  doi     = {10.1086/261816}
}

@article{azariadis1990threshold,
  author  = {Azariadis, Costas and Drazen, Allan},
  title   = {Threshold Externalities in Economic Development},
  journal = {Quarterly Journal of Economics},
  volume  = {105},
  number  = {2},
  pages   = {501--526},
  year    = {1990},
  doi     = {10.2307/2937797}
}

@article{durlauf1995multiple,
  author  = {Durlauf, Steven N. and Johnson, Paul A.},
  title   = {Multiple Regimes and Cross-Country Growth Behaviour},
  journal = {Journal of Applied Econometrics},
  volume  = {10},
  number  = {4},
  pages   = {365--384},
  year    = {1995},
  doi     = {10.1002/jae.3950100404}
}

@article{BJFP+20,
  title   = {Complex Economic Activities Concentrate in Large Cities},
  author  = {Balland, Pierre-Alexandre and Jara-Figueroa, Cristian and Petralia, Sergio G. and Steijn, Mathieu P. A. and Rigby, David L. and Hidalgo, C{\'e}sar A.},
  journal = {Nature Human Behaviour},
  volume  = {4},
  number  = {3},
  pages   = {248--254},
  year    = {2020},
  doi     = {10.1038/s41562-019-0803-3}
}

@article{krugman1991increasing,
  title={Increasing returns and economic geography},
  author={Krugman, Paul},
  journal={Journal of political economy},
  volume={99},
  number={3},
  pages={483--499},
  year={1991},
  publisher={The University of Chicago Press}
}

@book{fujita2001spatial,
  title={The spatial economy: Cities, regions, and international trade},
  author={Fujita, Masahisa and Krugman, Paul R and Venables, Anthony},
  year={2001},
  publisher={MIT press}
}

@incollection{duranton2004micro,
  title={Micro-foundations of urban agglomeration economies},
  author={Duranton, Gilles and Puga, Diego},
  booktitle={Handbook of regional and urban economics},
  volume={4},
  pages={2063--2117},
  year={2004},
  publisher={Elsevier}
}

@article{glaeser1992growth,
  title={Growth in cities},
  author={Glaeser, Edward L and Kallal, Hedi D and Scheinkman, Jose A and Shleifer, Andrei},
  journal={Journal of political economy},
  volume={100},
  number={6},
  pages={1126--1152},
  year={1992},
  publisher={The University of Chicago Press}
}

@incollection{rosenthal2004evidence,
  title={Evidence on the nature and sources of agglomeration economies},
  author={Rosenthal, Stuart S and Strange, William C},
  booktitle={Handbook of regional and urban economics},
  volume={4},
  pages={2119--2171},
  year={2004},
  publisher={Elsevier}
}

@article{fujita2002agglomeration,
  title={Agglomeration and market interaction},
  author={Fujita, Masahisa and Thisse, Jacques-Fran{\c{c}}ois},
  journal={Available at SSRN 315966},
  year={2002}
}

@article{Chen2022,
  author  = {Chen, Jiandong and Gao, Ming and Cheng, Shulei and Hou, Wenxuan and Song, Malin and Liu, Xin and Liu, Yu},
  title   = {Global 1~km $\times$ 1~km gridded revised real gross domestic product and electricity consumption during 1992–2019 based on calibrated nighttime light data},
  journal = {Scientific Data},
  year    = {2022},
  volume  = {9},
  number  = {1},
  pages   = {202},
  doi     = {10.1038/s41597-022-01322-5},
  note    = {Dataset DOI: 10.6084/m9.figshare.17004523}
}

@misc{rwi_hdx,
  author       = {{Center for International Earth Science Information Network (CIESIN) and Facebook Connectivity Lab}},
  title        = {Relative Wealth Index (RWI)},
  howpublished = {Humanitarian Data Exchange},
  year         = {2025},
  note         = {\url{https://data.humdata.org/dataset/relative-wealth-index} (accessed January 14, 2026)}
}

@article{dijkstra2019eu,
  title={The EU-OECD definition of a functional urban area},
  author={Dijkstra, Lewis and Poelman, Hugo and Veneri, Paolo},
  year={2019},
  publisher={OECD}
}

@article{tolbert1996us,
  title={US commuting zones and labor market areas: A 1990 update},
  author={Tolbert, Charles M and Sizer, Molly},
  year={1996}
}

@article{ghs_fua_2023,
  author    = {Florczyk, Aneta J. and Corbane, Christina and Ehrlich, Daniele and Freire, Sérgio and Pesaresi, Martino and Schiavina, Marcello and Kerle, Norman and others},
  title     = {GHSL Data Package 2023: GHS Functional Urban Areas (GHS-FUA)},
  year      = {2023},
  publisher = {Joint Research Centre (JRC), European Commission},
  doi       = {10.2760/70029},
  url       = {https://ghsl.jrc.ec.europa.eu/}
}

@article{barca2012case,
  title={The case for regional development intervention: place-based versus place-neutral approaches},
  author={Barca, Fabrizio and McCann, Philip and Rodr{\'\i}guez-Pose, Andr{\'e}s},
  journal={Journal of regional science},
  volume={52},
  number={1},
  pages={134--152},
  year={2012},
  publisher={Wiley Online Library}
}

@article{patel2021new,
  title={The new era of unconditional convergence},
  author={Patel, Dev and Sandefur, Justin and Subramanian, Arvind},
  journal={Journal of Development Economics},
  volume={152},
  pages={102687},
  year={2021},
  publisher={Elsevier}
}

@article{alvaredo2018elephant,
  author  = {Alvaredo, Facundo and Chancel, Lucas and Piketty, Thomas and Saez, Emmanuel and Zucman, Gabriel},
  title   = {The Elephant Curve of Global Inequality and Growth},
  journal = {AEA Papers and Proceedings},
  year    = {2018},
  volume  = {108},
  pages   = {103--108},
  doi     = {10.1257/pandp.20181073}
}

@article{anselin2022spatial,
  title={Spatial econometrics},
  author={Anselin, Luc},
  journal={Handbook of spatial analysis in the social sciences},
  pages={101--122},
  year={2022},
  publisher={Edward Elgar Publishing}
}

@article{moreno2021metropolitan,
  title={Metropolitan areas in the world. Delineation and population trends},
  author={Moreno-Monroy, Ana I and Schiavina, Marcello and Veneri, Paolo},
  journal={Journal of Urban Economics},
  volume={125},
  pages={103242},
  year={2021},
  publisher={Elsevier}
}

@book{milanovic2016global,
  title={Global inequality: A new approach for the age of globalization},
  author={Milanovic, Branko},
  year={2016},
  publisher={Harvard University Press}
}

@article{redding2017quantitative,
  title={Quantitative spatial economics},
  author={Redding, Stephen J and Rossi-Hansberg, Esteban},
  journal={Annual Review of Economics},
  volume={9},
  number={1},
  pages={21--58},
  year={2017},
  publisher={Annual Reviews}
}

@incollection{combes2015empirics,
  title={The empirics of agglomeration economies},
  author={Combes, Pierre-Philippe and Gobillon, Laurent},
  booktitle={Handbook of regional and urban economics},
  volume={5},
  pages={247--348},
  year={2015},
  publisher={Elsevier}
}

@article{olaberria2023reversal,
  title={The reversal problem: Development going backwards},
  author={Olaberria, Eduardo and Reinhart, Carmen M},
  year={2023},
  publisher={Banco Central de Chile}
}

@misc{patel2025convergence,
  author       = {Patel, Dev and Sandefur, Justin and Subramanian, Arvind},
  title        = {We Were Wrong about Convergence},
  year         = {2025},
  month        = dec,
  howpublished = {ChatGDP Blog},
  url          = {https://www.chat-gdp.org/we-were-wrong-about-convergence/},
  note         = {Accessed: 2026-01-08}
}

@article{gennaioli2014growth,
  title={Growth in regions},
  author={Gennaioli, Nicola and La Porta, Rafael and Lopez De Silanes, Florencio and Shleifer, Andrei},
  journal={Journal of Economic growth},
  volume={19},
  number={3},
  pages={259--309},
  year={2014},
  publisher={Springer}
}

@article{bartkowska2012regional,
  title={Regional convergence clubs in Europe: Identification and conditioning factors},
  author={Bartkowska, Monika and Riedl, Aleksandra},
  journal={Economic Modelling},
  volume={29},
  number={1},
  pages={22--31},
  year={2012},
  publisher={Elsevier}
}

@article{mellander2015night,
  title={Night-time light data: A good proxy measure for economic activity?},
  author={Mellander, Charlotta and Lobo, Jos{\'e} and Stolarick, Kevin and Matheson, Zara},
  journal={PloS one},
  volume={10},
  number={10},
  pages={e0139779},
  year={2015},
  publisher={Public Library of Science San Francisco, CA USA}
}

@article{perez2021night,
  title={Are night-time lights a good proxy of economic activity in rural areas in middle and low-income countries? Examining the empirical evidence from Colombia},
  author={P{\'e}rez-Sind{\'\i}n, Xaqu{\'\i}n S and Chen, Tzu-Hsin Karen and Prishchepov, Alexander V},
  journal={Remote Sensing Applications: Society and Environment},
  volume={24},
  pages={100647},
  year={2021},
  publisher={Elsevier}
}

@article{bickenbach2016night,
  title={Night lights and regional GDP},
  author={Bickenbach, Frank and Bode, Eckhardt and Nunnenkamp, Peter and S{\"o}der, Mareike},
  journal={Review of World Economics},
  volume={152},
  number={2},
  pages={425--447},
  year={2016},
  publisher={Springer}
}

@article{ahrend2014makes,
  title={What makes cities more productive? Evidence on the role of urban governance from five OECD countries},
  author={Ahrend, Rudiger and Farchy, Emily and Kaplanis, Ioannis and Lembcke, Alexander C},
  year={2014},
  publisher={OECD}
}

@article{musso2025large,
  title={Large cities lose their growth advantage as urban systems mature},
  author={Musso, Andrea and Rybski, Diego and Helbing, Dirk and Neffke, Frank},
  journal={arXiv preprint arXiv:2510.12417},
  year={2025}
}

@misc{un_inequality,
  author       = {{UN75}},
  title        = {Inequality: Bridging the Divide},
  year         = {n.d.},
  howpublished = {\url{https://www.un.org/en/un75/inequality-bridging-divide}},
  note         = {Accessed: 2025-05-23}
}
